\documentclass[conference]{IEEEtran}
\IEEEoverridecommandlockouts
\usepackage[ruled,vlined]{algorithm2e}
\usepackage{amsmath}
\usepackage{xspace}
\usepackage[binary-units=true]{siunitx}
\usepackage{ulem}
\usepackage{xcolor}
\usepackage{graphicx}
\usepackage{hyperref}
\usepackage{tabularx}

\hypersetup{colorlinks=true,linkcolor=black,citecolor=blue,filecolor=black,urlcolor=blue}

\def\BibTeX{{\rm B\kern-.05em{\sc i\kern-.025em b}\kern-.08em
    T\kern-.1667em\lower.7ex\hbox{E}\kern-.125emX}}
    
\begin{document}

\title{The benefits of prefetching for large-scale cloud-based neuroimaging analysis workflows}

\newcommand{\tristan}[1]{\color{orange}\textbf{From Tristan: }#1\color{black}}
\newcommand{\tristanmod}[2]{\color{orange}\sout{#1}{#2}\color{black}}

\newcommand{\ariel}[1]{\color{blue}\textbf{From Ariel:}#1\color{black}}
\newcommand{\arielmod}[2]{\color{blue}\sout{#1}{#2}\color{black}}

\newcommand{\valerie}[1]{\color{purple}\textbf{Valerie: }#1\color{black}}
\newcommand{\valeriemod}[2]{\color{purple}\sout{#1}{#2}\color{black}}

\newcommand{\aws}{AWS\xspace}
\newcommand{\hcp}{HCP\xspace}
\newcommand{\sfs}{S3Fs\xspace}

\author{Valerie Hayot-Sasson$^1$, Tristan Glatard$^1$, Ariel Rokem$^2$ \\ $^1$ Department of Computer-Science and Software Engineering, Concordia University, Montreal, Canada\\ $^2$ Department of Psychology and eScience Institute, University of Washington, Seattle, Washington, USA}

\maketitle
\begin{abstract}
To support the growing demands of neuroscience applications, researchers are transitioning to cloud computing for its scalable, robust and elastic infrastructure. Nevertheless, large datasets residing in object stores may result in significant data transfer overheads during workflow execution.
Prefetching, a method to mitigate the cost of reading in mixed workloads, masks data transfer costs within processing time of prior tasks. We present an implementation of ``Rolling Prefetch", a Python library that implements a particular form of prefetching from AWS S3 object store, and we quantify its benefits.

Rolling Prefetch extends \sfs, a Python library exposing AWS S3 functionality via a file object, to add prefetch capabilities. 
In measured analysis performance of a 500~GB brain connectivity dataset stored on S3, we found that prefetching provides significant speed-ups of up to 1.86$\times$, even in applications consisting entirely of data loading. The observed speed-up values are consistent with our theoretical analysis. Our results demonstrate the usefulness of prefetching for scientific data processing on cloud infrastructures and provide an implementation applicable to various application domains.

\end{abstract}


\section{Introduction}


Many fields of science are experiencing increases in the volume of datasets available to researchers. Neuroscience in particular is experiencing a rapid growth in data due to technical advances, 
scientific breakthroughs, and sociological trends towards open data sharing. This increase in data is providing the basis for new discoveries about brain structure and function, but it also presents technical challenges. To deal with the deluge of available data, neuroscientists are increasingly adopting cloud platforms for data storage and processing. However, inefficient use of cloud services can lead
to needlessly longer processing time and cost. We aim to investigate
methods to reduce the impact of data transfers for
neuroimaging workflows on cloud services.

Data prefetching
is a well-established technique for the reduction of data access-related costs~\cite{callahan1991software, mowry1992design,
klaiber1991architecture}. Traditionally, prefetching was
used to reduce memory latency, as memory accesses were significantly slower than CPU processing. However, since the rise of Big Data,
prefetching has also been shown to be beneficial to the processing of large
datasets located on remote storage~\cite{yildiz2018improving}. During the
execution of an application, data required for future tasks are
copied from the remote storage device to compute-local storage, such that
when the application requires the data, it can read it from local
storage.

A recent example of the effectiveness of prefetching on the cloud is Netco~\cite{jalaparti2018netco}, a prefetching extension
integrated into the Hadoop Distributed File System (HDFS). Future data is prefetched based on two measures: 1) size of the
data to be processed and 2) task deadline.  Netco demonstrated superior performance compared to other file systems, which importantly motivates our study. However, it remains tightly bound to HDFS while cloud applications generally use different file systems. A more versatile solution is needed that would broadly apply to cloud data analysis and storage.

The present study focuses on neuroscience data that describe long-range
connections between different parts of the human brain, a central research topic in contemporary
neuroscience~\cite{bassett_network_2017}. The three-dimensional trajectory
of the major neural pathways, composed of millions of neuronal axons, are
inferred from measurements of diffusion MRI and processed using computational tractography algorithms. These
algorithms generate ``streamlines": 3D curves that approximate the
trajectories of the major pathways. A single human brain measurement may
contain several millions of these streamlines, with their coordinates
assessed at sub-millimeter resolution. Subsequent analyses of these
streamlines usually access
streamlines sequentially and entirely within the files in which they are stored. Such an access pattern creates an excellent opportunity for prefetching.

This paper investigates the benefits of prefetching for cloud-based processing of neuroscience streamlines. Through
both theoretical analysis and experimentation, we characterize the
speed-up provided by prefetching compared to sequential data transfers for
neuroscience data processing deployed on the Amazon Web Services cloud. More specifically, this paper makes the following contributions: 
\begin{itemize}
    \item Formalization and performance analysis of a ``rolling prefetch" data  scheme for cloud-based applications;
    \item Implementation based on the S3Fs Python library to access data on Amazon S3;
    \item Experimental evaluation in the Amazon cloud with a 500-GB dataset of streamlines derived from dMRI data.
\end{itemize}


\section{Materials and Methods}

Our implementation of prefetching, known as Rolling Prefetch, is a Python library implemented as a layer on top of \sfs. The code is available under the MIT license at \url{https://github.com/ValHayot/rollingprefetch}.
\sfs is a Python library, based on FSSpec, for interacting with directories and files located on Amazon S3.
To offset the cost of latency on S3, \sfs leverages the on-demand caching mechanisms provided by FSSpec.

\begin{figure}
\begin{center}
\includegraphics[width=\columnwidth]{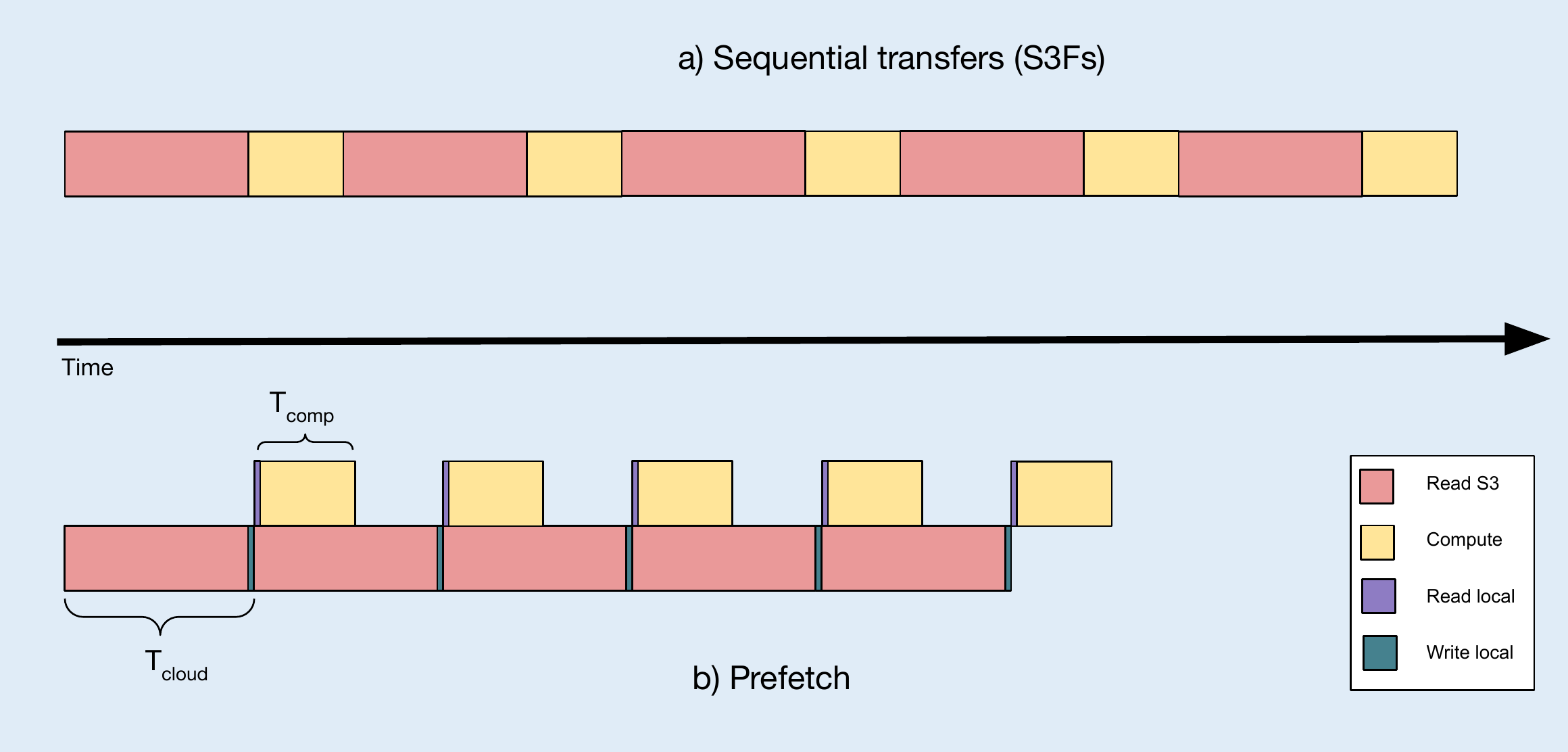}
\end{center}
\caption{Prefetching vs sequential transfers (\sfs)}
\label{fig:prefetch}
\end{figure}

Unlike \sfs which has distinct data transfer and compute phases, prefetched data transfers happen concurrently with compute (Figure~\ref{fig:prefetch}). In both cases, data is transferred from cloud storage in blocks of configurable size. While local I/O overheads occur during prefetch, the resulting overhead is in general minimal since memory disks are expected to be used for this purpose. Most importantly, prefetching requires knowledge of the application's data access pattern, such that transfers can be correctly anticipated. In our case, we assume that data is read sequentially, as is commonly the case with tractography data.

\subsection{Algorithm}


Rolling Prefetch combines prefetching with data eviction. By evicting data post application processing, it ensures a reduced footprint on local storage. Rolling Prefetch consists of three threads: (1) the reading thread loads data blocks from local storage and marks them for eviction, (2) the prefetching thread transfers data blocks from cloud storage to local storage, and (3) the eviction threads deletes the blocks that have been marked for eviction.
    
    
    
        
        
        
    

\subsubsection{Reading}

The reading thread is triggered by the application to read data directly from cache or to wait until it is prefetched, if not found in the cache. 
By waiting for the data to be cached, we ensure that performance is comparable to \sfs in a worst case scenario. Furthermore, as all data reads
are considered to be sequential, if the current block is not in memory then it is the current block being prefetched. Whenever a prefetched block has been read fully, it is up to the
read function to flag it for deletion.

\subsubsection{Prefetching}

Throughout the lifetime of the application, the prefetching thread continuously requests blocks from cloud storage, so long as there remain blocks that have not been prefetched (Algorithm~\ref{alg:prefetch}). Each block will be written to an appropriate cache location while not exceeding user-defined space limits.

Initially, the \texttt{used} variable for all cache locations is set to zero and the \texttt{file\_index} variable
is set to point to the first file index in the sequential ordering. The algorithm then enters a while loop whose
termination is dictated by the main program.
If all files have been prefetched, the prefetching threads terminates regardless of the status of the main thread.
Within the loop, the algorithm iterates between all available cache locations in priority order. The amount of available storage space
remaining is computed, and if it is insufficient to write the next block, the algorithm iterates through the file list $total\_files$
($verify\_used()$) and queries the filesystem to determine which blocks have been evicted. Based on how many blocks have been
removed, we can update the amount of used space, and by extension, the amount of available space. Should sufficient space
be released, the next block is fetched from cloud storage and is written to local storage. If there is insufficient space, the
next local storage location is tried.


\begin{algorithm}
\SetAlgoLined
    \SetKwInOut{Input}{Input}
    \Input{
    $fetch$ a shared variable indicating whether the main thread has terminated or not\\
    $total\_files$ the list of all the files to prefetch;\\
    $cache\_location$ list of paths to cache locations;\\
    $total$ total space available on prefetch cache;\\
    $total\_size$ cumulative sum of all prefetched files;\\
    $blocksize$ size of a block\\}
    
$used \gets 0$\;
$file\_index \gets 0$\;
\While{$fetch$}{
\ForEach{$cache\_location$}{
$available \leftarrow total - used$\;
\If{ $available < blocksize$}{
    $available \gets verify\_used()$\;
}
\If{ $available \geq blocksize$}{
    $fetch\_next\_block()$\;
    $used \leftarrow used + blocksize$\;
}
}
\If{$total\_size \leq offset$ or $file\_index + 1 < total\_files$}{
    $file\_index \gets file\_index + 1$\;
    $total\_size \gets sizeof(total\_files[file\_index])$\;
}
\ElseIf{$total\_size \leq offset$}{
    break;
}
}
\caption{Prefetching}\label{alg:prefetch}
\end{algorithm}

\subsubsection{Eviction}

Similar to prefetching, the eviction thread only terminates when the main thread has completed processing. 
To avoid additional costs caused by continuously querying the filesystem to determine which files can be evicted,
we determine the names of all the blocks that should be evicted ($get\_all\_blocks$), and verify whether they exist
in the filesystem at time of removal. We then update the list to ensure that we do not attempt to remove this file more 
than once. Between each loop, we sleep for 5 seconds to ensure that sufficient time has elapsed between evictions.

The eviction thread ensures deletion of all remaining files prior to terminating.

    
    
                
                


\subsection{Performance analysis}

We consider the cost of processing a file using sequential transfers (\sfs) to be the sum of three components (Equation~\ref{eq:s3fs}):
1) latency-associated costs, 2) bandwidth-associated costs and 3) application compute time.
The cost of latency is experienced every time a chunk of data is requested from cloud storage. That is, if we read a full file in a single call to cloud storage, we will only pay for latency once. If we issue
multiple calls to obtain file data on cloud storage, we will pay latency for each call. In contrast,
bandwidth is not affected by the number of calls. Whether we request all the data at once
or in chunks, bandwidth remains fixed, and time to transfer the data is
the quotient between the total amount of data transferred and the bandwidth. Compute time is assumed to be proportional to data size, as is frequently the case in neuroimaging.

    \begin{equation}
T_{\mathrm{seq}} = n_{b}l_c + \frac{f}{b_{cr}} +  cf,
\label{eq:s3fs}
\end{equation}
where $n_{b}$ is the number of data blocks, $f$ is the total size of the file to transfer, $l_c$ is the cloud latency, $b_{cr}$ is the cloud read bandwidth, and $c$ is the compute time per byte consumed.

Rolling Prefetch contrasts Equation~\ref{eq:s3fs} in that the compute and data transfer times mask one another (Equation~\ref{eq:prefetch}). However, Rolling Prefetch has a slightly higher performance penalty than that of sequential transfers when there is no compute, due to reading and writing the data to local storage. Furthermore, we
must consider the initial read from cloud storage, where no compute can occur concurrently, and the last compute, where no data transfer can occur concurrently.

\begin{equation}
T_{\mathrm{pf}} = T_{\mathrm{cloud}} + (n_b-1)\max\left(T_{\mathrm{cloud}}, T_{\mathrm{comp}}\right) + T_{\mathrm{comp}} \label{eq:prefetch}
\end{equation}
where $T_{\mathrm{cloud}}$ is the time to download a block from the cloud and write it to local storage, and $T_{\mathrm{comp}}$ is the time to read a block from local storage and process it:
\begin{eqnarray*}
T_{\mathrm{cloud}} &=& \overbrace{l_{c} + \frac{f}{b_{cr}n_{b}}}^{\mathrm{cloud\ read}} + \overbrace{l_{l} + \frac{f}{b_{lw}n_{b}}}^{\mathrm{local\ write}}, \\
T_{\mathrm{comp}} &=& \underbrace{l_l+\frac{f}{b_{lr}n_b}}_{\mathrm{local\ read}} + \underbrace{\frac{cf}{n_b}}_{\mathrm{compute}},
\end{eqnarray*}
where $l_l$ is the latency of local storage, $b_{lw}$ is the write bandwidth to local storage, and $b_{lr}$ is the read bandwidth to local storage.

This simple model provides several insights. First, if we neglect local transfers ($l_l$=0, $b_{lw}$=$b_{lr}$=+$\infty$), a reasonable assumption when local storage is in-memory, then we have:
\begin{equation*}
T_{\mathrm{seq}} = T_{\mathrm{pf}} + (n_b-1)\min\left(T_{\mathrm{cloud}},T_{\mathrm{comp}}\right),
\end{equation*}
and therefore the speed-up provided by Rolling Prefetch compared to sequential transfers is:
\begin{equation}
S=\frac{T_{\mathrm{seq}}}{T_{\mathrm{pf}}}=1+(n_b-1)\frac{\min\left(T_{\mathrm{cloud}},T_{\mathrm{comp}}\right)}{T_{\mathrm{pf}}} < 2 
\label{eq:speed-up}
\end{equation}
Rolling Prefetch is therefore expected to provide a speed-up of at most 2x. This upper bound is approached when $T_{\mathrm{cloud}} \approx T_{\mathrm{comp}}$, which requires that cloud transfer and compute times are of similar magnitude.

In sequential transfers, using a single block ($n_b=1$) leads to the shortest transfer time. In practice, 
block size is of course constrained by available memory. In contrast, with Rolling Prefetch, an optimal block size $\hat n_b$ exists under the reasonable assumption that $l_l \ll l_c$:
\begin{equation}
    \hat n_b = \sqrt{\frac{cf}{l_c}}
\end{equation}
It suggests that on a given cloud infrastructure, the number of blocks should increase when the application compute time or the total data size increases.

Finally, as the number of blocks increases, $T_{\mathrm{seq}}$ and $T_{\mathrm{pf}}$ become equivalent to $n_bl_c$ and $n_b(l_c+l_l)$, respectively, resulting in parallel asymptote lines.

\subsection{Data}

To demonstrate the utility of Rolling Prefetch in neuroimaging, we analyzed data derived from a dMRI experiment conducted in a single subject. These data are special, in that a  Super-Resolution Hybrid Diffusion Imaging (HYDI) method was used to achieve an effective spatial resolution of 0.625 mm$^3$, much higher than the typical $\sim$2 mm$^3$ commonly used. The details of the HYDI method and parameters of data acquisition were previously described \cite{Elsaid2019-ez}. The source MRI data were 14.94 GB. The data were processed using a probabilistic tractography algorithm \cite{Berman2008-xg}, implemented as part of DIPY \cite{Garyfallidis2014-el} and accelerated using CUDA to work on GPU \cite{rokem2021gpu}. This  generated $\sim$498 GB of streamline data. Tractography streamlines are usually stored in the neuroscience-specific \texttt{.trk} file format. In this case, streamlines were sharded into 464 \texttt{.trk} files stored in an S3 bucket using the high-availability Standard storage class. 

 Each \texttt{.trk} file is comprised of a 1,000-byte header and a body of variable length consisting 
of a series of streamlines. Each streamline section contains \SI{4}{\byte} that denote the number of points in the streamline,
followed by a series of floating point values detailing each coordinate and ends with a series of values representing properties of the streamline. We used Nibabel, a Python library that reads and writes  neuroscience-specific data formats, to read these files. For \texttt{.trk} files, Nibabel can return individual
data points via a generator. This is known as ``lazy loading'' and can be turned on or off prior to reading the file. As a result of the data representation format in \texttt{.trk} files, Nibabel reads may incur significant overhead: because it issues a total of three read calls for each streamline in the file.
In addition, an affine transform is stored together with the data, and used to bring coordinates from different measurements into register. For example, when Nibabel reads \texttt{.trk} files it automatically applies an affine transformation to the coordinates of each streamline stored in the file. This means that some amount of compute is always executed when data is read from file.

\subsection{Experiments}
To evaluate the performance Rolling Prefetch, we conducted 4 experiments: 1) varying
the number of files; 2) varying the blocksize; 3) parallel processing of files;  and 4) neuroimaging use-cases. For experiments 1-3, the pipeline consisted of reading the file with Nibabel. As Nibabel
performs a minimal amount of computing, Rolling Prefetch is believed to have an effect even here. Additional compute required
by real analysis should only increase the benefit of using prefetching.

\subsubsection{Varying the number of files}\label{exp:files}

We vary the number of \texttt{.trk} files read to determine when Rolling Prefetch will be favourable relative to \sfs. We expected that for small amount of data, there should be no significant discernible difference between \sfs and Rolling Prefetch. This is because compute time is
expected to be minimal and the blocksize large compared to the total dataset size. In other words, there
will be less opportunities to prefetch data with very small datasets, unless the block size is proportionally smaller,
in which case latency will be penalizing both reads from \sfs and Rolling Prefetch, extending processing time. With larger files, there will be more computation occurring during the reads, and therefore, more opportunities
to mask the data transfers within the compute.

We benchmarked the time it took to lazily read 1, 5, 10, 15, 20 and 25 files and extract the streamlines in Nibabel.
The total data size for each of the file increments are 1.1, 5.8, 11.9, 18.0, 24.2 and \SI{31.2}{\gibi\byte},
respectively. The blocksize was set to \SI{64}{\mebi\byte} and the prefetch storage was located on \texttt{tmpfs} with
a limit of \SI{2}{\gibi\byte} of storage space. We performed 10 repetitions.

\subsubsection{Varying the blocksize}\label{exp:blocksize}
In this experiment, we aimed to quantify the effect of the number of blocks ($n_b$) on Rolling Prefetch and \sfs. According to our previous analysis, we expect that \sfs will reach its best performance with a single block, and that the runtime will linearly increase with $n_b$ from that point on. For Rolling Prefetch, we expect the runtime to decrease to a minimum, and then increase linearly with $n_b$.

Block sizes of 8, 16, 32, 64, 128, 256, 512, 1024,
and \SI{2048}{\mebi\byte} were used to load the files into Nibabel and extract the streamlines.
We use the 
largest-sized block (\SI{2048}{\mebi\byte}) to determine the
overhead of using the Rolling Prefetch algorithm, as no file in the HYDI dataset is larger than \SI{1.7}{\gibi\byte},
and therefore, there are no opportunities to prefetch.

Five files pertaining to the HYDI dataset were used to generate the results.
These files ranged from \SI{793}{\mebi\byte} to
\SI{1.5}{\gibi\byte} in size each. Since only Rolling Prefetch is capable of treating a list of files as a single file, a new
Nibabel object needed to be created for each file in the case of \sfs.

The experiment was executed on both \sfs and Rolling Prefetch for each block and repeated 10 times. Tmpfs was excluded from this
experiment as the \sfs blocksize would have no effect on its performance. Prefetched storage was set to be tmpfs and
configured to have \SI{2}{\gibi\byte} of available space such that the largest block size could fit in prefetch
storage.

\subsubsection{Parallel processing}\label{exp:parallel}

Since perceived
throughput may be affected by the number of threads used, we aimed to determine if the performance difference
between \sfs and Rolling Prefetch remains proportional to the amount of data being processed per thread. 

If throughput is reduced by the number of active threads, it is expected that 
the cost of data transfers will outweigh the cost of computing, such that the difference between \sfs and Rolling Prefetch will
be very small. Furthermore, since S3 is a scalable distributed file system and local storage may not be (or may be more limited in terms of scalability), local storage
throughput is expected to be reduced significantly with the addition of multiple threads. 

We used a total of four parallel processes to evaluate the effects of parallel processing. We increased the number of
files per thread between each condition from 1-20, with the last condition loading
a total of \SI{108}{\gibi\byte} into Nibabel. The blocksize was set to \SI{64}{\mebi\byte} and we set the prefetch
storage location to be on \texttt{tmpfs}, with a limit of \SI{1}{\gibi\byte} of storage per process. 10 repetitions were
performed.

\subsubsection{Neuroimaging use cases}\label{exp:4}
Our previous experiments evaluate the value of Rolling Prefetch when the application performs the most
basic and necessary action: converting the data from binary to array representation.
We expect that the benefits of Rolling Prefetch can be maximized with an increased compute time, due to greater opportunities to prefetch data during compute. However, we also expect that there is a 
peak ratio between data transfer time and compute time (data-to-compute ratio), where the benefits of Rolling Prefetch will be highest.
If compute time is significantly greater than data transfer, regardless of all the time saved by prefetching data, the
total runtime will predominantly be compute. This motivates further profiling to determine how to shard
large datasets, like the HYDI dataset, such that the benefits of Rolling Prefetch can be maximized.

Since the benefits of Rolling Prefetch may vary according to the ratio between data transfer and compute time, we have selected two use cases with varying computation times: 1) histogram distribution of streamline lengths; and 2) bundle recognition.

The distribution of streamline lengths is used in order to assess the performance of tractography algorithms. It may also aid in determining if the files are good candidates for compression. The histogram computation consists of loading each streamline within the dataset lazily, gathering the lengths of
each individual streamline and generating a histogram of 20 bins from their lengths.

The human brain is composed of multiple different functional regions, and these regions are connected through major anatomical pathways. One of the first steps in the analysis of streamline data from human brain dMRI data is to classify the streamlines according to their three-dimensional trajectory into these different pathways. This is referred to as ``bundle recognition"~\cite{garyfallidis2018recognition}. To demonstrate the benefits of Rolling Prefetch in this use-case, we ran a software pipeline that determines whether streamlines within the tractography file belong to either of two different bundles (the corticospinal tract, CST and the arcuate fasciculus, ARC), or to neither \cite{Kruper2021-bo}. Currently, the pipeline loads the data all at once (i.e. no lazy loading) and then performs the bundle recognition task. Thus, it is not possible to read only fragments of a large file which does not fit in memory, in this case, and as a consequence of the separation of data loading and processing, there will be no opportunities to keep the prefetched data transfers masked within compute operations. 

We chose not to modify the pipeline, in order to determine what is the speedup provided
by Rolling Prefetch without additional optimization, and in order to be able to speculate on the possible benefits of adapting existing pipelines
to support larger-than-memory files and allowing reads and computations to occur together.

Varying parameters were used for the bundle recognition and
histogram application. For instance, recognition was
executed on two different compute instances (c5.9xlarge and r5.4xlarge). The experiment executed on the c5.9xlarge instance consisted of a \SI{1}{\gibi\byte} file with a \SI{64}{\mebi\byte} blocksize. The histogram experiment executed on the r5.4xlarge instance used 10 files totalling \SI{12}{\gibi\byte} and used a \SI{32}{\mebi\byte} blocksize. 

We also considered the speedups obtained from
bundle recognition in smaller files. We split up the 
previous \SI{1}{\gibi\byte} file and split it up into
9 shards containing the same number of streamlines. File sizes ranged from \SI{73}{\gibi\byte} to \SI{165}{\mebi\byte}. We compared the speedup provided by Rolling Prefetch on a \SI{165}{\mebi\byte} shard to
the processing of all 9 shards. These experiments were
run exclusively on the r5.4xlarge instance with 10 repetitions.

To compare the benefits of Rolling Prefetch with different data-to-compute ratios, we executed the histogram computation and the bundle recognition algorithm on
an r5.4xlarge instance using the \SI{1}{\gibi\byte} file. Blocksize in both cases was fixed at \SI{32}{\mebi\byte}. 5 repetitions were performed.

In all cases, \SI{2}{\gibi\byte} of cache storage was allocated to Rolling Prefetch.

\subsection{Infrastructure}

For the first three experiments, we used an Amazon EC2 t2.xlarge instance with 4 vCPUs and \SI{16}{\gibi\byte} of RAM with a
maximum of \SI{7.8}{\gibi\byte} of RAM of tmpfs space. The compute instance was configured to use Red Hat Enterprise
Linux 8.3 with kernel version 4.18.0. This instance was hosted in availability zone \texttt{us-west-2a}.
The experiments all used Python version 3.9.0 with version 0.5.2 of \sfs and Nibabel version 3.2.1.

The scientific applications required significantly more memory than available on t2.xlarge instances. As a result, the scientific applications were executed on a c5.9xlarge instance and an r5.4xlarge configured identically
to the t2.xlarge instance, but located in the \texttt{us-west-2d} availability zone. c5.9xlarge instances consist of 64 vCPUs and have \SI{72}{\gibi\byte} of available memory, whereas the r5.4xlarge instance consists of 16 vCPUs and \SI{128}{\gibi\byte} of available memory. 

The data used for all experiments originated from the HYDI tractography dataset stored on the Amazon S3 object store. 
This dataset is broken down into 464 files ranging from \SI{700}{\mebi\byte} to \SI{1.7}{\gibi\byte} each. This dataset
was also located in the \texttt{us-west-2} region.

To get an idea of the cost of data transfers, we measured the time it would take to read data sequentially from S3 to the t2.xLarge instance. We varied the filesize from \SI{1}{\kibi\byte} to \SI{4}{\gibi\byte} and measure the time it took to read the files from
a \texttt{us-west-2} S3 bucket and from tmpfs. This method was favoured over more standard benchmarking libraries to ensure the
two filesystems were benchmarked in the same way. Results were then compared to benchmarking results produced by
the \href{https://github.com/dvassallo/s3-benchmark}{s3-benchmark} library to ensure validity of our benchmarking
method. The results can be seen in Table~\ref{table:benchmarks}.

\begin{table}
\caption{Measured latency and bandwidth obtained from reading files ranging from \SI{1}{\kibi\byte} to \SI{4}{\gibi\byte} on a t2.xLarge instance reading from an S3 bucket in the same region.}
\centering
\begin{tabular}{| c | c | c| }
\cline{2-3}
  \multicolumn{1}{c|}{}& S3 & memory \\ 
  \hline
 bandwidth (MB/s) & 91 & 2221 \\  
 latency (s) & 0.1 & $\num{1.6e-6}$ \\
 \hline
\end{tabular}

\label{table:benchmarks}
\end{table}

\section{Results}
\subsection{Varying the file size}
As we increase the file size we observe that the disparity between \sfs and
Rolling Prefetch increases (Figure~\ref{fig:filesize}), with Rolling Prefetch significantly outpacing ($\sim$1.7x faster) 
\sfs at \SI{31.2}{\gibi\byte} (25 files). This is expected as per our theoretical evaluation. With a large amount of blocks, the pipeline duration is determined by the maximum between the time it takes to
compute the data $T_{comp}$ and the time it takes to transfer the data $T_{cloud}$. On the other hand, \sfs runtime is the sum of the time it takes to transfer the data and the compute. In this case, the compute time is large enough such that using Rolling Prefetch significantly reduces runtime with large amounts of data.

As the size of the data decreases, the benefit of using Rolling Prefetch also decreases. 
In this particular use case, S3 data transfer time is much larger
than that of compute time, thus the only time we can save is that of $T_{comp}$. $T_{comp}$ like ${T_{cloud}}$ is
proportional to the number of blocks used, and thus the amount of time saved increases with file size. In the 
worst case, Rolling Prefetch performance equals that of \sfs.

\subsection{Varying the block size}

\begin{figure}
\begin{center}
\includegraphics[width=\columnwidth]{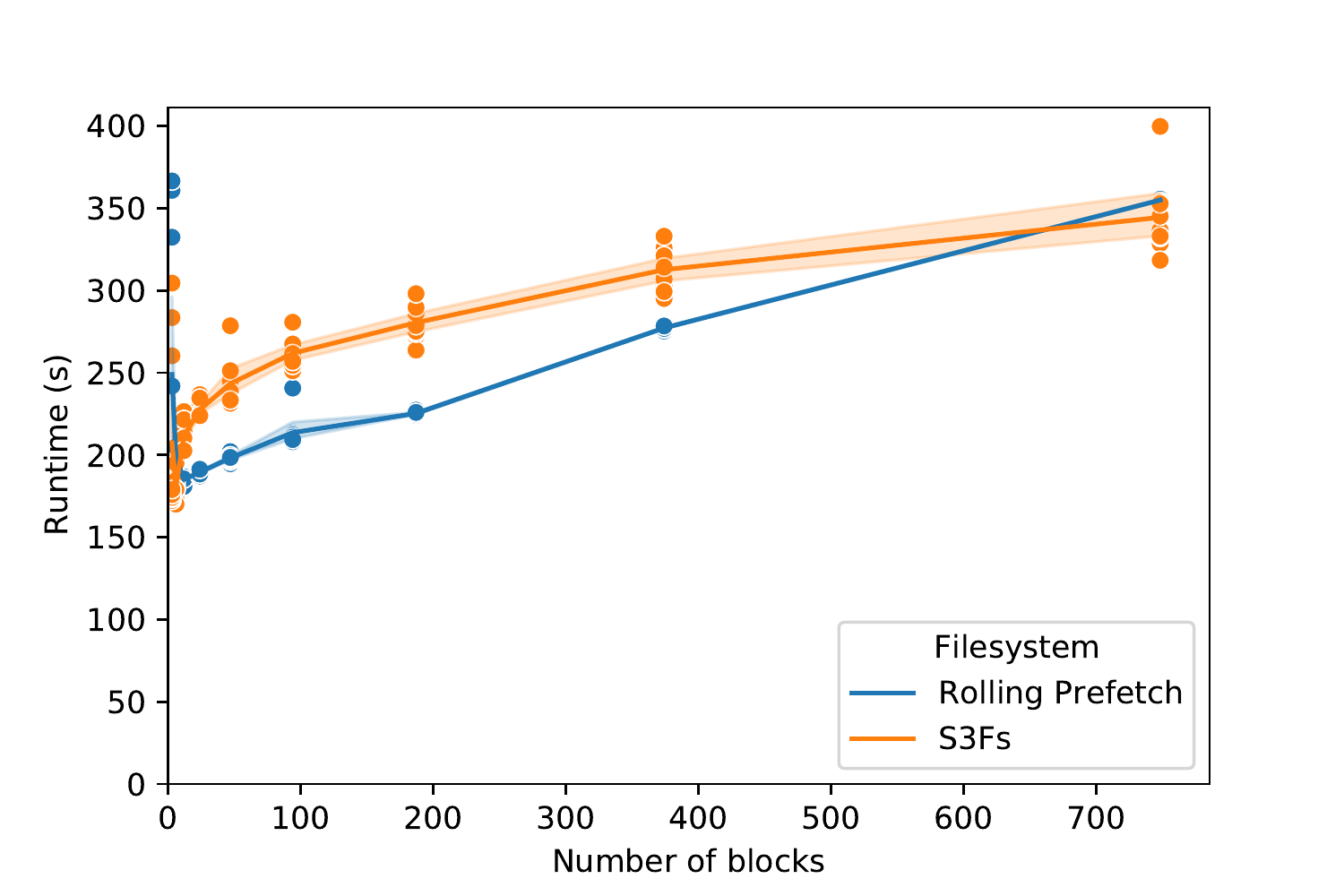}
\setlength{\abovecaptionskip}{-10pt}
 \setlength{\belowcaptionskip}{-20pt}
\caption{Runtime performance of reading a \SI{6}{\gibi\byte} subset of the HYDI dataset stored on Amazon S3 into Nibabel using
\sfs and Rolling Prefetch with various block size configurations on an Amazon EC2 t2.xlarge instance. Prefetch cache consisted of  \SI{2}{\gibi\byte} tmpfs storage.}
\label{fig:blocksize}
\end{center}
\end{figure}

At a very large number of blocks, the performance of both \sfs and Rolling Prefetch is degraded due to
increased impacts of both S3 storage latency and additionally, in the case of Rolling Prefetch, local storage latency (Figure~\ref{fig:blocksize}).  \sfs starts to surpass the speed of Rolling Prefetch, although not substantially (approx. 1.03~X) at 748 blocks.  Furthermore, while we read from S3
at certain blocksizes, the block size on local storage is dictated by filesystem and kernel readahead 
size. In our specific case, the data would be read from tmpfs, and thus we do not see a significant added cost
to Rolling Prefetch. It could be inferred that the overhead of Rolling Prefetch may be more significant with a large
number of blocks due to Nibabel's access patterns, which consist of many small reads.

Conversely, when decreasing the number of blocks we start to see an increase in performance in both \sfs and Rolling Prefetch. This
is because both benefit from decreased latency to access data on S3, by generally making fewer read calls to it. Rolling Prefetch is found to be faster than \sfs with a peak speedup of 1.24~X at 187 blocks (i.e. 
\SI{32}{\mebi\byte} blocks). The speedup obtained from Rolling Prefetch did not vary significantly between 24 and
187 blocks, averaging at a speedup of approximately 1.2~X. Rolling Prefetch outpaces \sfs at less than
748 blocks as the cost of local latency has diminished sufficiently with the larger blocks. As the blocksize
increases, Rolling Prefetch is able to begin masking S3 latency and bandwidth within its compute.
\sfs and Rolling Prefetch performance converges again when the blocks reach a certain size, placing more weight disk bandwidth
than latency. \sfs exceeds the performance of Rolling Prefetch at a single block, where no actual prefetching can take
place and we pay the cost of implementation overheads, such as writing the data to local storage 
and reading it from local storage rather than directly from memory.

Generally speaking, variations in blocksize did not lead to major speedups between Rolling Prefetch and \sfs. Since the
total files size and compute time was fixed, there was a fixed upper bound to how many data transfers could
be masked by compute time. By minimizing the amount of latency through a reduction in the number of blocks, the 
overall application time was shortened, in both \sfs and Rolling Prefetch, and thus a larger percentage of the total 
runtime could be masked by compute.

\subsection{Parallelized workloads}
\begin{figure}
\begin{center}
\includegraphics[width=\columnwidth]{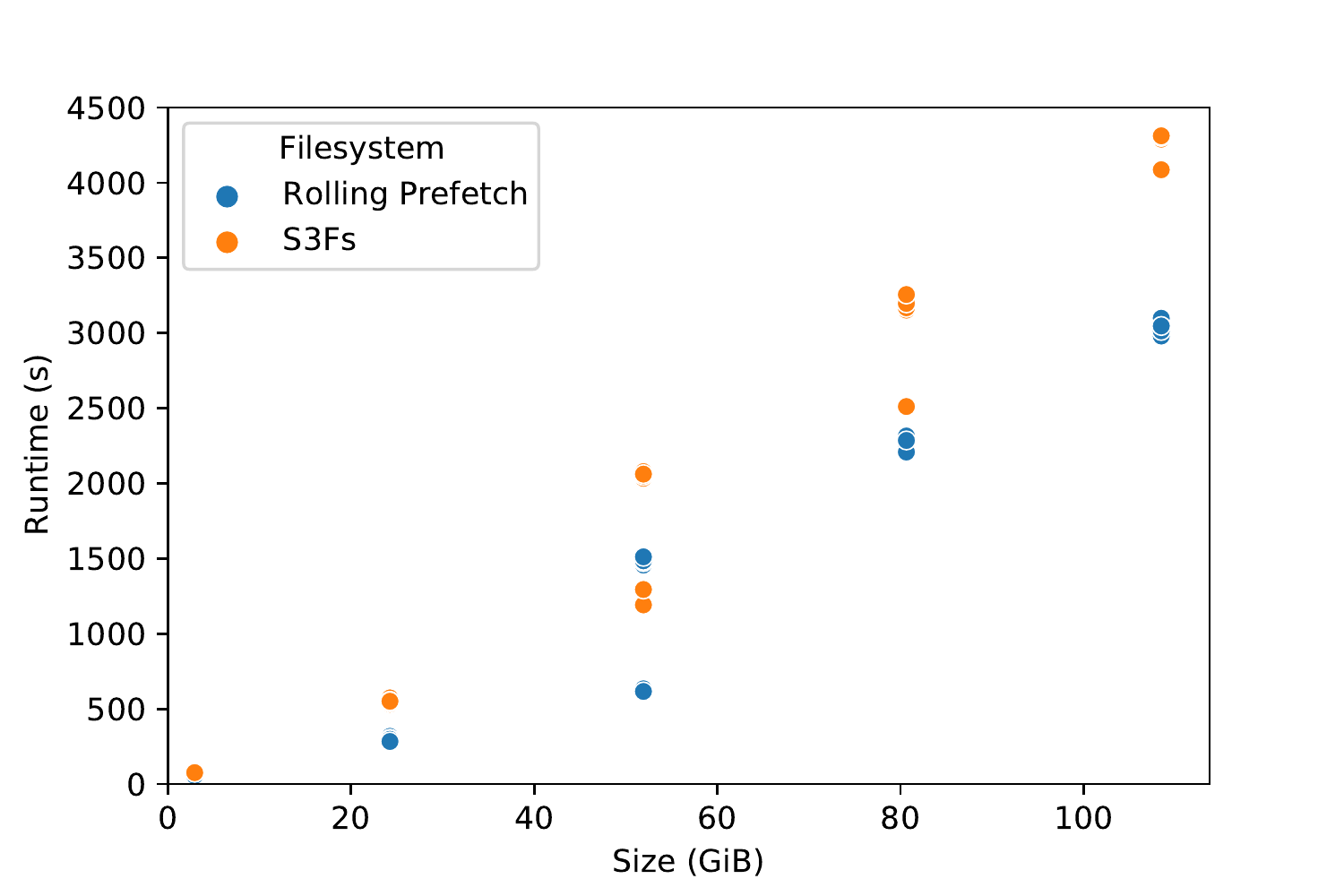}
\setlength{\abovecaptionskip}{-10pt}
\setlength{\belowcaptionskip}{-10pt}
\caption{Runtime performance of reading subsets of the HYDI dataset stored on Amazon S3 into Nibabel in parallel using 4
\sfs and Rolling Prefetch processes. Blocksize was set to \SI{64}{\mebi\byte} on both \sfs and Rolling Prefetch.
Prefetch cache consisted of \SI{1}{\gibi\byte} tmpfs storage.}
\label{fig:parallel}
\end{center}
\end{figure}

When we parallelize the reads up to four concurrent processes (Figure~\ref{fig:parallel}), we notice that the trends are still consistent
despite increased contention on local filesystems. This is likely due to the fact that \texttt{tmpfs} speeds
would remain sufficiently high even with the increased contention caused by having at minimum 4 and 
potentially even up to 8 concurrent threads accessing the filesystem at the same time. We speculate that the same 
pipeline running with data read from a single disk would have had worse performance
due to its lack of scalability.

The maximum speedup achieved, on average, during this experiment was 1.86$\times$ at 24.2~GiB. The average speedup
was around 1.52~x altogether, with the minimum average speedup being 1.37$\times$ at 80.6~GB. Due to the high
variability in results within both Rolling Prefetch and \sfs, we expect that these speedup differences may be a
result of bandwidth and latency variability across runs. We do not expect that conditions are more favourable
for reading the data with four threads processing 24.2~GiB; since they are processing the
same amount of data concurrently, the overall speedup should be constant.

\subsection{Neuroimaging use-cases}

\begin{figure}
\begin{center}
\includegraphics[width=\columnwidth]{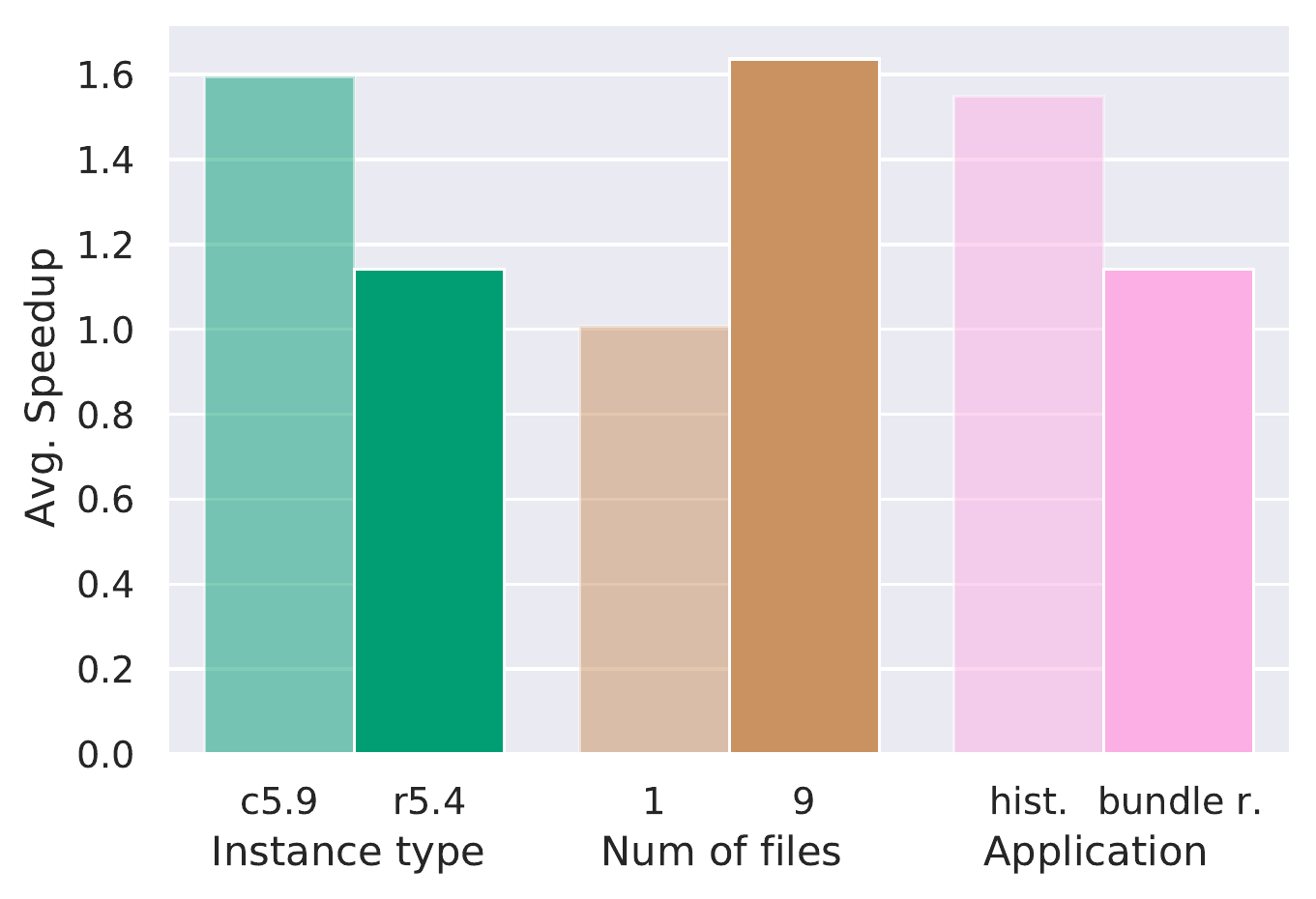}
\setlength{\abovecaptionskip}{-10pt}
\setlength{\belowcaptionskip}{-8pt}
\caption{Rolling prefetch speedup of the neuroimaging use-cases (histogram and bundle recognition) in various conditions. Experiments varying instance type and number of files were only performed with the bundle recognition pipeline.} 
\label{fig:speedup}
\end{center}
\end{figure}

The results obtained from our neuroimaging use cases indicate that, in all
cases, the use of Rolling Prefetch can speed up analyses (Figure~\ref{fig:speedup}).
The extent of this speedup is, however, dependent on a number of
factors, such as instance type, application and number of shards.

\begin{figure}
\begin{center}
\includegraphics[height=160pt]{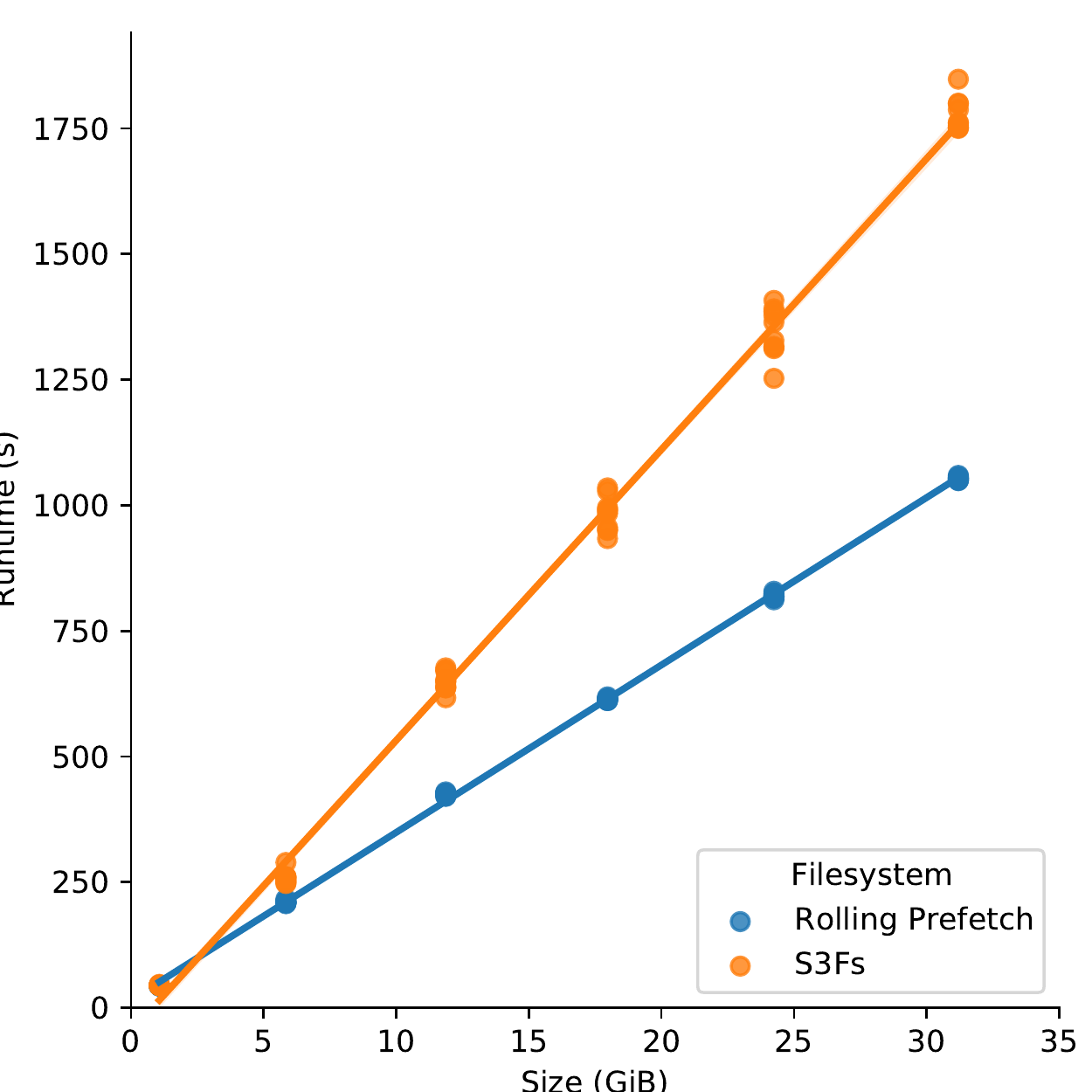}
\setlength{\abovecaptionskip}{0pt}
\setlength{\belowcaptionskip}{-10pt}
\caption{Runtime performance of reading subsets of the HYDI dataset stored on Amazon S3 into Nibabel using \sfs and Rolling Prefetch on an Amazon EC2 t2.xlarge instance. Blocksize was set to \SI{64}{\mebi\byte} on both \sfs and
Rolling Prefetch. Prefetch cache consisted of \SI{2}{\gibi\byte} tmpfs storage.}
\label{fig:filesize}
\end{center}
\end{figure}

The greatest observed speedup with Rolling Prefetch was found when executing the bundle recognition application
with a greater number of shards (1.64$\times$ faster). This result echoes what was observed in Figure~\ref{fig:filesize} with even a more compute-intensive application like 
bundle recognition. We do not, however, observe a speedup with a single shard, as the size of the shard was so small it did not incur many reads from S3.

Due to the compute-intensive nature of the bundle recognition pipeline, it comes as no
surprise that overall speedup in the unsharded case is minimal (1.14$\times$). The ratio of data-to-compute was found to be approximately 1/7, where the compute took, on average around 9000s. Thus, even if Rolling Prefetch significantly speeds up reads, the data transfer time is
minimal to the overall compute time of the application. With a highly data-intensive
workflow, such as a histogram computation, we observe that the speedup is more
significant with a speedup of 1.5$\times$. Although our model dictates that the upper bound
of speedup that can be obtained with Rolling Prefetch is 2$\times$, the observed speedups
obtained by these two applications never reach that boundary. A possible explanation may be that the data-to-compute ratio of histogram computation was too large to mask many of
the reads within the compute, whereas the data-to-compute ratio was too small in the segmentation application, such that any speedup obtained
through the Rolling Prefetch algorithm was offset by the computation time. The ideal
algorithm for Rolling Prefetch would lie somewhere between the data and computational
requirements of the histogram and segmentation applications.

Although the execution parameters differed between the two instance types thus not
making them directly comparable, we observed a greater speedup on the c5.9xlarge
(1.6$\times$) than on the r5.4xlarge (1.1$\times$) instance. We suspect that the speedup obtained
on the c5.9xlarge instance was relatively high as a result of the increased parallelism
obtained from the increased number of CPUs, resulting in a more data-intensive application. While the r5.4xlarge instance did process a larger number of files, 
suggesting that speedup should be greater, the actual file sizes varied, and with
it, the streamline features, potentially resulting in a more compute-intensive
execution further exacerbated by the decrease in available vCPUs.



\section{Discussion}

Our theoretical analysis and experimental results demonstrate that there is a substantial processing time gain to be
obtained from using Rolling Prefetch, particularly in the case of mixed workloads, where there is a significant
cost associated with time spent on compute and data transfers. This works well with typical use cases in neuroimaging,
where tasks vary widely ranging from very short tasks to long ones necessitating hours to complete and where datasets
are large enough to incur transfers of similar magnitudes. Moreover, to save time with data transfers, researchers
may opt to transfer their data over the compute instance, and perform the computation exclusively with
data stored locally. While this does effectively give the optimal performance during processing, researchers are left to manage
the data themselves. Since local storage on compute can become quite costly, researchers must decide between
processing only a subset of the data, paying hefty fees related to storing large amounts of data or 
incorporating data management logic directly into their workflows. 

There are also natural limitations to
Rolling Prefetch. For instance, in the case of parallel workloads, in certain instances S3 would be
preferred to the available local storage. This is a consequence of the fact that S3 is designed to be scalable
and is capable of processing many requests simultaneously, whereas unless configured to do so, local storage
will not scale adequately to increased contention.

\subsection{Benefits of Rolling Prefetch}

Rolling Prefetch is an added layer on top of \sfs that replaces its built-in cache strategies to intercept
application read calls and ensure that data is preloaded. The implementation ensures that filesystem space requirements are not exceeded during prefetching through limits set by the user. 
With the built-in eviction mechanism, users are not required to do any form of data management should they be processing datasets larger than available storage. Furthermore, the library allows configuration of multiple storage device as cache for prefetching. Currently,
files are written to specific storage devices based on available space and storage priority in the list.

With just a simple implementation performed on a computation based on reading, we have observed up to a 1.8$\times$ speedup and a maximum overhead of 1.03$\times$. These observed speedups were
obtained when we set the cache location to a \texttt{tmpfs} device and may naturally decrease with a slower device such as an SSD. In our particular case, the speedups meant saving 20~min of processing time
on loading nearly \SI{100}{\gibi\byte} of data with a maximum runtime of approximately 71~min. Moreover, this
was achieved on a small instance with only \SI{1}{\gibi\byte} of prefetch storage allotted to each parallel
process, indicating that we can achieve good speedups even when resources are constrained. 

\subsection{Parallelism-related overheads}

While our experiments demonstrate that Rolling Prefetch continues to provide a performance improvement to parallel
applications running on S3, we do expect  performance to decrease if we continue to scale horizontally on
a single instance, or use a slower device as cache. Our implementation consists of two threads actively reading and writing to local storage. Each time the number of Rolling
Prefetch instances increase, we double the amount of threads writing to local storage. Although it is standard
to have multiple DIMMs on a single machine, it may
not be necessarily true of other storage devices. That being said, attached cloud volumes may
also be sufficiently scalable such that processing time remains unaffected by an increase in processes.

To reduce the load of prefetching data to local storage, we can add a third component to Rolling Prefetch that periodically tracks cache bandwidth. Using such a parameter, the algorithm could take into consideration filesystem 
load in addition to user-specified priority.

\subsection{Task granularity}
\sfs was designed to be used within distributed processing frameworks. With such frameworks, we must take into consideration task granularity and how that would affect
scheduling. Rolling prefetch becomes beneficial with larger files. Assuming a workload where the task is simply to load the streamlines, we would need a few
GBs of data to start noticing a speedup. In cases where resources are ample enough to run all tasks in
parallel and the amount of data processed by a task is minimal, Rolling Prefetch and \sfs would probably behave
similarly, with \sfs potentially performing a bit faster.

The risk with passing large amounts data to the library within a task is that it is less robust to failures as any failed task would have to resume from the beginning. This could
take a significant amount of time, depending on how much data had been processed by the task prior to failure.
It is understandable that implementations such as Netco are adaptations to
persistent filesystem, since then prefetching could work without limiting scheduler logic. For our 
implementation to be efficient within parallel processing frameworks, we would need to decouple it from
the task itself and allow it to communicate with the task scheduler to determine which files to prefetch and
where.

\section{Conclusion}

Overall, we conclude that Rolling Prefetch would be a valuable addition to large-scale neuroimaging pipeline executing
on the cloud, particularly in instances where data transfer and compute time are similar. 
Future work should focus on the following improvements: The loading part of Rolling Prefetch should be performed outside of task themselves to ensure that it does not interrupt any form of task scheduling and to avoid tasks which take too long to restart if lost. 
The library should also make sure to be able to communicate with the schedulers to help determine where tasks
should be scheduled give the location of the data.

\section{Acknowledgments} 

This research was supported by NIH grant 1RF1MH121868-01 and by cloud compute credits from Microsoft Azure. This research was also supported by the Canada Research Chairs program, and by the Natural Sciences and Engineering Research Council of Canada.

\bibliographystyle{plain}
\bibliography{biblio}

\end{document}